# Dynamic Critical Behavior in Models of Ferromagnetic Gadolinium

A. K. Murtazaev and V. A. Mutailamov*

*Institute of Physics, Dagestan Scientific Center, Russian Academy of Sciences, Makhachkala, 367003 Russia*

*\* e-mail: vadim.mut@mail.ru*

A numerical technique combining Monte Carlo and molecular dynamics simulations is used for the first time to examine the complex critical dynamics of models of ferromagnetic gadolinium in which both strong exchange interactions and relativistic effects of several different types are taken into account. A finitesize scaling technique is used to calculate the corresponding dynamic critical exponents. The role played by isotropic dipole–dipole interaction in the critical behavior of gadolinium is evaluated. The results obtained provide an explanation for the anomalous dynamic critical behavior of gadolinium.

## 1. INTRODUCTION

Analysis of dynamic critical properties is a major problem in statistical physics and theory of phase transitions [1–3]. The substantial progress achieved to date in this field became possible mainly through theoretical and experimental studies. Nevertheless, the development of a rigorous and consistent theory of dynamic critical phenomena based on microscopic Hamiltonians remains a challenging problem in the modern theory of phase transitions and critical phenomena [1, 4].

Current knowledge of critical dynamics has mostly been gained in mode-coupling and dynamic scaling theories [1–3], which were developed independently on the basis of essentially different ideas. However, many predictions of these theories are in good agreement. In [3], universality classes of dynamic critical behavior were proposed in the framework of the dynamic scaling theory. It was shown that the dynamic universality classes depend not only on the space dimension $d$, the number $n$ of order-parameter components, the interaction range, and the symmetry of the Hamiltonian, but also on the energy and order-parameter conservation laws [3]. The classification has played an important role in understanding critical dynamics and is being used to this day. The underlying dynamic scaling hypothesis provides an adequate characterization of dynamic critical behavior. However, the classification is not flawless: its basic ideas are not supported by solid evidence, even though the ensuing predictions are consistent with many, but not all, experimental findings. Moreover, the classification is not complete.

The critical dynamics of magnetically ordered crystals, especially ferromagnets, is characterized by a great diversity and complexity due to the importance of both exchange interaction and relatively weak relativistic effects (such as anisotropy and dipole–dipole interaction). The most essential factor of the latter kind is dipole–dipole interaction, which plays an increasingly important role as the critical point is approached. Note that the aforementioned classification of universal dynamic critical behavior does not allow for any effect due to dipole–dipole interaction. Further analyses (see [5–7]) have shown that theories taking into account dipole–dipole interaction predict dynamics of two types, normal and stiff, each characterized by a specific set of critical exponents.

There is no conclusive experimental evidence, because the available data are mutually inconsistent [4]. Experimental validation is further complicated by the fact that exchange, anisotropy, and dipole–dipole interaction can contribute to the behavior of a real material simultaneously, in which case critical dynamics obviously depend on their relative strengths. Moreover, since there exist crossover regions near the critical point, the actual critical behavior may change as the critical point is approached. The real behavior due to relativistic effects is even more diverse: in particular, anisotropy can be uniaxial, cubic, etc., while dipole– dipole interaction can be either isotropic or anisotropic. Simultaneous influence of these factors may partially explain the inconsistency of experimental data on the dynamic critical properties of magnetically ordered materials.

It is clear that experimental studies can hardly elucidate the current discrepancy between observations and predictions, because high-precision measurement of critical parameters is a very difficult task. Moreover, since almost any experimental result is due to the combined effects of all factors, their individual strengths and contributions are practically impossible to single out. The problem is also unlikely to be amenable to rigorous theoretical analysis in view of enormous mathematical difficulties.

In recent studies, methods of computational physics have played an increasingly important role in dealing with complex issues of this kind. Highly accurate and reliable calculations of critical parameters can be performed by applying these methods, at least, in studies of static critical behavior [8]. In particular, the advantages of Monte Carlo and molecular dynamics simulations include not only rigorous mathematical foundations and error control within their respective frameworks, but also the possibility of evaluating the relative importance of individual parameters.

The key characteristics of critical dynamics are the critical exponent $w$ for the relaxation time $\tau$ and the dynamic critical exponent $z$:

$$\tau \infty |t|^{-w}, \; \tau \infty \xi^z,$$

where

$$t = |T - T_c|/T_C, \; \xi = (T/T_c - 1)^{-\nu}.$$

In the mid-1990s, a new technique was proposed for calculating the dynamic critical exponent $z$, based on dynamic finite-size scaling theory [9] and a special procedure for determining the characteristic frequency $\omega_c$.

In this paper, methods of computational physics are used to analyze critical dynamics in models of ferromagnetic gadolinium. The key questions to be answered here can be formulated as follows.

I. How does isotropic dipole–dipole interaction affect dynamic critical behavior?
II. Is there any difference in critical dynamics between different directions in noncubic crystals?
III. Can the computational technique employed in this study be used to expose the dependence of critical dynamics on weak factors, such as dipole–dipole interaction?

The present analysis of models of gadolinium is motivated by the following observations.

1. The static critical behavior of gadolinium is strongly affected by isotropic dipole–dipole interaction [13, 14].
2. The controversial results obtained in extensive experimental studies of the critical dynamics of ferromagnetic gadolinium [4, 15–18] do not provide sufficient basis for any final conclusion.
3. The complex dynamic critical behavior of gadolinium has been analyzed in a number of theoretical studies [19, 20].
4. The dynamic critical behavior of gadolinium is of interest per se, because it is determined by the combined effects of three factors: exchange interaction, magnetocrystalline anisotropy, and isotropic dipole–dipole interaction.
5. The static critical behavior of gadolinium has been thoroughly examined in both experimental [13] and numerical [14] studies, and their results provide a good basis for an analysis of critical dynamics.

Note that the technique employed in the present analysis of critical dynamics was previously used in our studies of the dynamic critical behavior of the classical Heisenberg model [21] and models of the multisublattice antiferromagnet $Cr_2O_3$ [21, 22].

## 2. MODEL

Gadolinium is a rare-earth metal with hexagonal close-packed lattice. The Curie temperature of the order–disorder second-order phase transition between ferromagnetic and paramagnetic states in gadolinium is $T_C$ = 293 K. In a previous study, we proposed models of ferromagnetic gadolinium that take into account all basic properties of this materials [14] and examined their static behavior. A model Hamiltonian for gadolinium can be represented as follows [14]:

$$H = -\frac{1}{2}\sum_{i,j} J(\vec{S}_i \vec{S}_j) - D\sum_i (S_i^z)^2 - D_{dip}\sum_i \langle \vec{M} \rangle \vec{S}_i, \quad |\vec{S}_i| = 1. \quad (1)$$

Here, the first term represents exchange interaction between a gadolinium ion and its twelve nearest neighbors, with $J > 0$; the second term, easy-axis anisotropy along the hexagonal axis, characterized by an anisotropy constant $D$; the third one, isotropic dipole–dipole interaction with coupling constant $D_{dip}$; $M$ is magnetization. We considered two models: in model I, only exchange interaction and anisotropy were taken into account; in model II, the dipole–dipole interaction was also included.

According to laboratory experiments [13], $D/J = 1.41 \times 10^{-4}$ and $D_{dip}/J = 1.35 \times 10^{-2}$. The coordinate system was defined so that the $z$ axis was aligned with the anisotropy axis, i.e., with the hexagonal axis of the crystal. The values of critical temperature were obtained in Monte Carlo simulations of the static critical behavior of these models [14].

## 3. TECHNIQUE

Our approach relies on dynamic scaling theory [1] and calculation of space- and time-displaced spin–spin correlation functions

$$C^k(\vec{r}_{12}, t) = \langle S_{\vec{r}_1}^k(t) S_{\vec{r}_2}^k(0) \rangle - \langle S_{\vec{r}_1}^k(t) \rangle \langle S_{\vec{r}_2}^k(0) \rangle, \quad (2)$$

where $\vec{r}_{12} = \vec{r}_1 - \vec{r}_2$, $S_{\vec{r}_1}^k(t)$ ($k = x, y, z$) is the spin at the site $\vec{r}_1$ at an instant $t$, $S_{\vec{r}_2}^k(0)$ is the spin at the site $\vec{r}_2$ at the initial moment ($t = 0$), and angle brackets denote ensemble averaging. At $T \approx T_C$ and sufficiently long times, the second term on the right-hand side of (2) can be neglected [11].

The space-time Fourier transform of (2) defines the dynamic structure factor

$$S^k(\vec{q}, \omega) = \int d\vec{r} \int_{-\infty}^{+\infty} C^k(\vec{r}, t) e^{-i(\vec{q}\vec{r} - \omega t)} dt, \quad (3)$$

where $\vec{q}$ is wavevector and $\omega$ is frequency. In the general case, the dynamic structure factor is proportional to the directly measurable neutron scattering function, being shifted by a constant frequency [1].

In accordance with the dynamic scaling hypothesis, the characteristic frequency is the median frequency defined by the relation

$$\int_{-\omega_c(\vec{q},\xi)}^{+\omega_c(\vec{q},\xi)} S^k(\vec{q}, \omega) d\omega = \frac{1}{2} \int_{-\infty}^{+\infty} S^k(\vec{q}, \omega) d\omega. \quad (4)$$

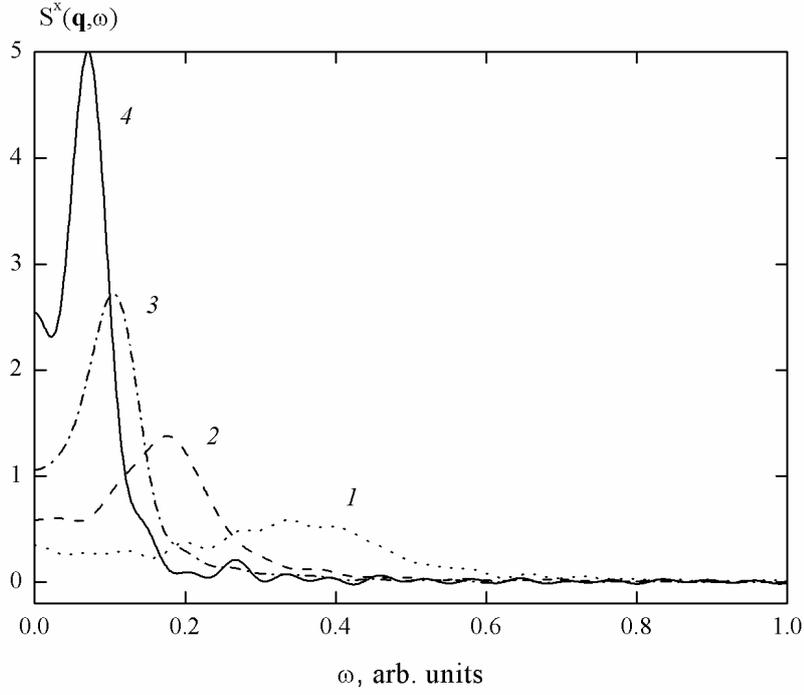

**Fig. 1.** Dynamic structure factor versus frequency in model I with $qL = 2\pi$ and $N = 432$ (*1*), 1024 (*2*), 2000 (*3*), and 3456 (*4*).

In the general case, the characteristic frequency depends on wavevector $\vec{q}$ and correlation length $\xi$. In dynamic scaling theory [1], it is postulated that

$$\omega_c(\vec{q},\xi) = q^z \Omega(q\xi), \quad (5)$$

where $\Omega$ is an unknown homogeneous function of $q\xi$.

In the models considered here, expression (3) for a system of size $L$ simulated at the critical point over a finite time interval $t_{\text{cutoff}}$ is represented as [10]

$$S^k(\vec{q},\omega) = \frac{1}{\sqrt{2\pi}} \sum_{\vec{r}_1,\vec{r}_2} \exp[i\vec{q}(\vec{r}_1-\vec{r}_2)] \int_{-t_{\text{cutoff}}}^{+t_{\text{cutoff}}} \exp(i\omega t) C^k(\vec{r}_1-\vec{r}_2,t) dt, \quad (6)$$

and Eq. (5) becomes

$$\omega_c \sim L^{-z} \Omega'(qL), \quad (7)$$

where $q = 2\pi n/L$ ($n = \pm 1, \pm 1, \ldots, L$). The sum in (6) corresponds to an integral over the space discretized into a lattice. Relation (7) is used to evaluate $z$ in practical calculations. The function $\Omega'$ is not known, but it is known to depend only on the product of $q$ and $L$. Therefore, the dynamic critical exponent can be evaluated by keeping it constant while varying the lattice size.

In each case considered in this study, the wavevector was aligned with a certain crystallographic axis, averaging over spins lying in alternating planes perpendicular to this axis was performed, and the averaged spins were used to calculate spin–spin correlation functions. In the case of a simple cubic lattice, these are crystallographic planes. The geometry corresponding to the hexagonal lattice of gadolinium is more complicated.

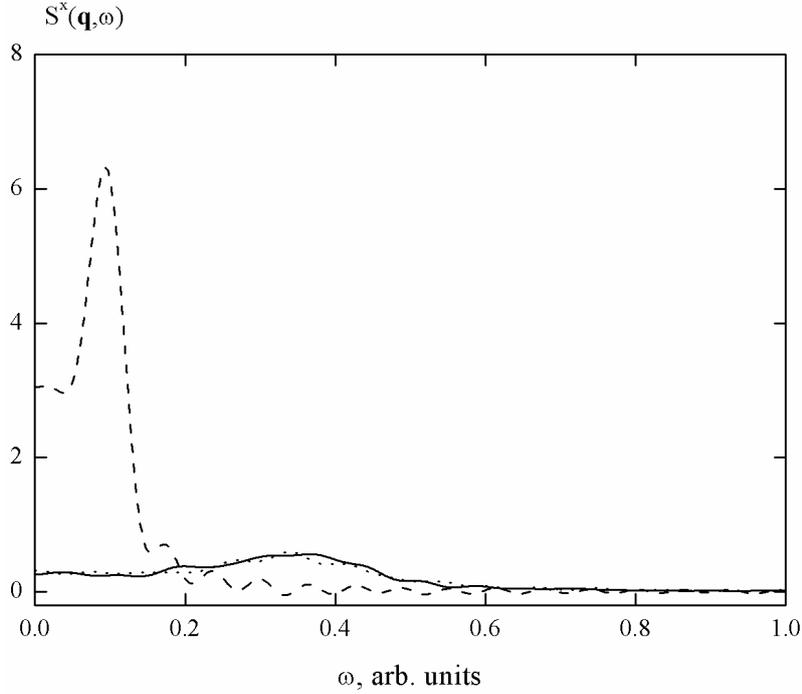

**Fig. 2.** Dynamic structure factor versus frequency in model I with $N = 432$ and $qL = 2\pi$ for wavevector aligned with the crystallographic axes $a$ (short-dash curve), $b$ (solid curve), and $c$ (long-dash curve).

The correlation functions were computed by using the system of differential equations of spin dynamics:

$$\frac{\partial \vec{S}_i}{\partial t} = [\vec{S}_i \times \vec{h}_{loc}], \quad t = t'\frac{g\mu_B}{J\gamma}, \quad |\vec{S}_i| = 1, \tag{8}$$

where $g$ is the Landé factor, $\gamma$ is the gyromagnetic ratio, and $\vec{h}_{loc}$ is the local field defined by (1). Before solving system (8), the standard Metropolis algorithm is executed to thermalize the system at the critical temperature [23].

We simulated systems with $L$ and number $N$ of magnetic sites varied from 4 to 18 and from 128 to 11 664, respectively, using periodic boundary conditions and calculating the three spin components for $\vec{q}$ aligned with three crystallographic axis. Note that the distances between the atomic planes are not equal when the wavevector is aligned with the $a$ or $b$ axis, whereas they are equal to $c/2$ in the case of alignment with the $c$ axis. The longest simulation time was $t_{cutoff} = 130$ (in arbitrary units), and the time step in computing the differential equations was $\Delta t = 0.01$. System (8) was solved by the fourth-order accurate Runge–Kutta method. Correlation functions (2) were computed by performing up to $n = 350$ averaging operations. It was found that the use of $\Delta t = 0.005$ and $n = 700$ did not lead to any significant change in numerical results.

The spin dynamics equations were solved by using several (up to ten) initial equilibrium configurations for each lattice size $L$, and the resulting characteristic frequencies $\omega_c$ were averaged.Correlation functions (2) were computed without using any normalization condition.

## 4. RESULTS

Figure 1 shows the dynamic structure factor $S^k(\vec{q},\omega)$ calculated as a function of frequency in model I for systems with different $N$ by using the $x$ components of spins in the case when the wavevector is aligned with the $a$ axis. The figure demonstrates that the maximum of the dynamic structure factor shifts toward smaller values of $\omega$ with increasing number of magnetic sites, while the absolute value of the maximum increases. The latter effect should be explained by the fact that correlation functions (2) were not normalized. However, the characteristic frequency is determined by the shape of the dynamic structure factor rather than its absolute value.

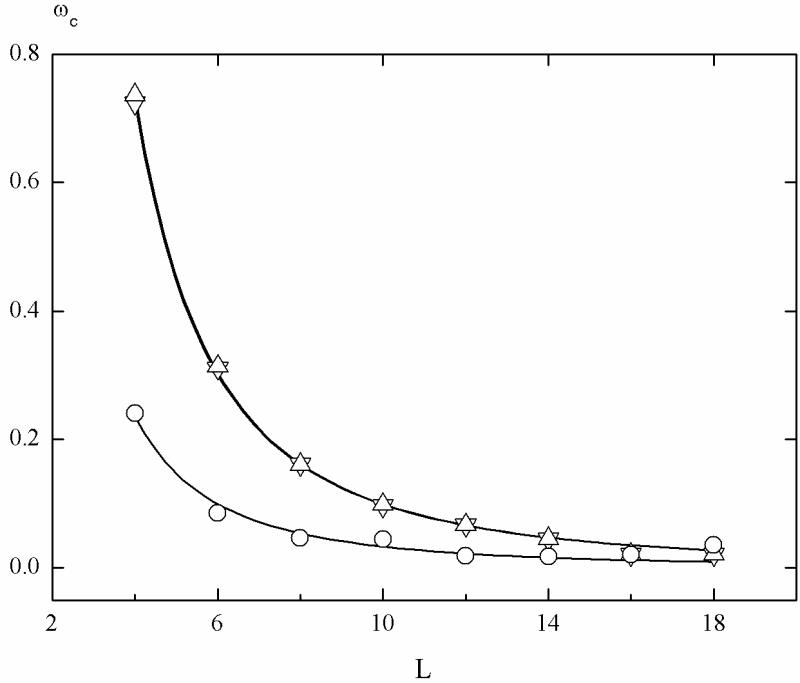

**Fig. 3.** Characteristic frequency versus lattice size in model I with $qL = 2\pi$ for wavevector aligned with the crystallographic axes $a$ ($\triangle$), $b$ ($\nabla$), and $c$ ($\circ$).

Figure 2 shows the dynamic structure factors calculated for different wavevector directions in model I by using the $z$ components of spins. Expectably, the shapes of the structure factors calculated for wavevectors aligned with the $a$ and $b$ axes are nearly identical and different from that calculated in the case when the wavevector is aligned with the hexagonal $c$ axis.

Figure 3 illustrates the dependence of $\omega c$ on $L$ in model I computed for different wavevector alignments by using the $x$ components of spins and averaging the characteristic frequency over several initial configurations. These results demonstrate similarity between the structure factors corresponding to characteristic frequencies $\omega_c$ for $\bar{q}$ aligned with the $a$ and $b$ axes and their substantial difference from those for $\bar{q}$ aligned with the $c$ axis.

The results obtained for model II are qualitatively similar to those presented in Figs. 1–3, but exhibit some quantitative differences.

The table summarizes the dynamic critical exponents $z$ evaluated by using the three spin components in both models and substituting the results illustrated by Fig. 3 into (7) for $\bar{q}$ aligned with three crystallographic axes.

In model I, the values of $z$ calculated by using the $x$ and $y$ spin components for $\bar{q} \parallel a$ and $\bar{q} \parallel b$ are in good agreement with theoretical predictions for anisotropic magnetic systems ($z = 2$, model $A$ in [3]). However, the values of the dynamic critical exponent determined by using the $z$ components for $\bar{q} \parallel a$ and for $\bar{q} \parallel b$ lie between those predicted theoretically for anisotropic magnetic systems ($z = 2$, model $A$) and isotropic ferromagnets ($z = 2.5$, model $J$ in [3]).

The values of $z$ calculated by using the $x$ and $y$ spin components in model II are higher by approximately 0.2, and those calculated by using the $z$ spin components are higher by amounts varying from 0.05 to 0.2.

Dynamic critical exponent *z* for three different wavevector directions and three
spin projections in models of ferromagnetic gadolinium

|         | $\vec{q} \parallel a$ | $\vec{q} \parallel b$ | $\vec{q} \parallel c$ |
|---------|-----------|-----------|-----------|
|         | Model I   |           |           |
| $k = x$ | 2.07±0.06 | 2.11±0.06 | 2.28±0.06 |
| $k = y$ | 2.06±0.06 | 2.08±0.06 | 2.36±0.06 |
| $k = z$ | 2.30±0.06 | 2.24±0.06 | 2.37±0.06 |
|         | Model II  |           |           |
| $k = x$ | 2.29±0.06 | 2.25±0.06 | 2.47±0.06 |
| $k = y$ | 2.26±0.06 | 2.27±0.06 | 2.49±0.06 |
| $k = z$ | 2.35±0.06 | 2.35±0.06 | 2.54±0.06 |

It is obvious that the difference between the values of *z* predicted in models I and II is due solely to the contribution of dipole–dipole interaction in model II, because all calculations were otherwise similar. Note also that all calculated critical exponents *z* have reasonable values.

It is obvious that some of our results are impossible to explain within the framework of the scheme proposed in [3], and so are some results obtained in experimental studies of critical dynamics of gadolinium [4, 15–20]. As noted in the Introduction, the classification of universal critical behavior proposed in [3] ignores certain additional factors, such as dipole–dipole interaction. Therefore, some new universality classes may have to be introduced. Even more complex behavior must be observed when several interactions of different types contribute simultaneously to critical dynamics, in addition to strong exchange interactions. This behavior is actually observed both in real materials and in our model II.

Furthermore, the dynamic critical behavior of a noncubic crystal must depend on direction, as suggested by the data listed in the table. This conjecture is corroborated by the results presented in [20], where it was demonstrated that longitudinal and transverse scaling functions analogous to Ω Ω in (5) exhibit totally different behavior depending on temperature (or crossover) region. Several crossover scenarios depending on temperature regime, anisotropy, and dipole–dipole interaction were described in [20] for Heisenberg models. In that study, the effects due to anisotropy and dipole– dipole interaction were treated separately because of enormous theoretical difficulties. In our model II, these effects are taken into account simultaneously.

We believe that many special characteristics exposed here, as well as the anomalous dynamic critical behavior of real gadolinium, can be explained by the combined effects of all factors. Note that a new dynamic universality class, model *J**, has been proposed for gadolinium in [19].

## 5. CONCLUSIONS

A numerical approach combining Monte Carlo and molecular dynamics simulations is used to examine the critical dynamics of real ferromagnetic gadolinium.

1. Dynamic critical behavior of noncubic crystals depends on the direction of wavevector relative to crystallographic axes.

2. Isotropic dipole–dipole interaction strongly contributes to the critical dynamics of magnetic systems.
3. The critical behavior of gadolinium that is difficult to explain theoretically should be attributed to the effects of anisotropy and dipole–dipole interaction combined with strong exchange interaction.
4. It is obvious that the developing application of methods of computational physics to dynamic critical phenomena not only leads to valuable results, but also reveals subtle effects, such as those due to dipole–dipole interaction.